\documentclass[12pt,showpacs,aps,prd]{revtex4}
\usepackage{graphicx}
\input epsf

\begin{document}
\title{Investigating different structures of the $Z_{b}(10610)$ and $Z_{b}(10650)$}
\author{Chun-Yu Cui, Yong-Lu Liu and Ming-Qiu Huang}
\affiliation{Department of Physics, National University of Defense
Technology, Hunan 410073, China}
\begin{abstract}
The recently observed narrow resonance $Z_{b}(10610)$ is examined with the assumptions both as a $B^{*}\bar{B}$ molecular state and a $[bd][\bar{b}\bar{u}]$ tetraquark state with quantum numbers $I^{G}J^{P}=1^{+}1^{+}$. Possible interpolating currents are constructed to describe the $Z_{b}(10650)$ as an axial-vector $B^{*}\bar{B}^{*}$ molecular state or a $[bd][\bar{b}\bar{u}]$ tetraquark state. Using QCD sum rules (QCDSR), we consider contributions up to dimension six in the operator product expansion (OPE) at the leading order in $\alpha_{s}$. The mass is obtained as $(10.44\pm0.23)~\mbox{GeV}$ for molecular state and $(10.50\pm0.19)~\mbox{GeV}$ for tetraquark state, both of which coincide with the $Z_{b}(10610)$. The results $m_{B^{*}\bar{B}^{*}}=(10.45\pm0.31)~\mbox{GeV}$ and $m_{[bd][\bar{b}\bar{u}]}=(10.48\pm0.33)~\mbox{GeV}$ are consistent with the $Z_{b}(10650)$.
\end{abstract}
\maketitle
\maketitle

\section{Introduction}
In the past years, the Babar, Belle, CLEO, D0, CDF and FOCUS collaborations reported many charmonium-like states which stimulate theorists' interests in revealing their underlying structures~\cite{Nora}. There comes to the consensus that these new states are not regular mesons or baryons. Various approaches have implied that a possible candidate is the exotic state, which means a complex structure such as a molecule, a tetraquark or a hybrid. This statement indicates that the counterpart may exist in the bottomonium region.

After the two charged bottomonium-like resonances $Z_{b}(10610)$ and $Z_{b}(10650)$ being observed in the $\pi^{\pm}\Upsilon(nS)~~(n = 1, 2, 3)$ and $\pi^{\pm}h_{b}(mP)~~(m = 1, 2)$ mass spectra by Belle Collaboration~\cite{10620}, many attempts have been made to investigate their possible configurations with various models~\cite{Bugg,Voloshin,nieves,Sun,Chen,Chiral,Guo,Zhang,Cleven,mehen,hosaka}, most of which support the $B^{*}{\bar B^{(*)}}$ molecular structure with $J^{P}=1^{+}$. In ref.~\cite{Bugg}, an explanation of two charged bottomonium-like resonances in terms of cusps at $B^*B^{(*)}$ channel is presented. In Ref.~\cite{Voloshin}, the authors discuss the special decay behaviour of the J=1 S-wave $B^*B^{(*)}$ molecular states and study radiative transitions from $\Upsilon(5S)$ to molecular bottomonium based on the heavy quark symmetry. In Ref.~\cite{Guo}, $Z_b(10610)$ and $Z_b(10650)$ are interpreted as tetraquark states in the framework of chromomagnetic interaction Hamiltonian model. In Ref.~\cite{Chiral}, the authors investigate the mass spectra of S-wave bound states $B\bar{B}^*$ and $B^*\bar{B}^*$ systems with quantum numbers $I(J^{PC})=1(1^{+-})$ in the framework of a chiral quark model. All in all, quantum numbers compatible with the experiment are the fundamental ingredients in theoretical analysis of composite particles within any specific model. We notice that $Z_{b}(10610)$ was observed in the $\Upsilon(5S)\rightarrow Z_{b}(10610)+\pi^{-}$ decay process. Since the $\Upsilon(5S)$ has negative G-parity, due to emission of the pion, the newly observed $Z_{b}(10610)$ has positive G-parity. In consideration that it is charged, the known quantum numbers of $Z_{b}(10610)$ are $I^{G}=1^{+}$. Angular distributions analysis favors the $J^{P}=1^{+}$ assignment for $Z_{b}(10610)$~\cite{10620}. The $Z_{b}(10610)$ considered as a B*B molecular state in Ref.~\cite{Zhang} has $J^{P}=1^{+}$ with no definite quantum numbers of isospin and G-conjugation. Considering all their known quantum numbers $I^{G}J^{P}=1^{+}1^{+}$, we construct a molecular state with a B and a B* mesons using the configuration $B^{-}B^{*0}+B^{*-}B^{0}$.
It is difficult to construct a suitable axial-vector style molecular current interpolating the state $Z_{b}(10650)$ using both $B^{*}$ and $\bar{B^{*}}$ fields. A possible interpolator is supposed to describe the axial-vector style molecular states $B^{*}\bar{B^{*}}$:
\begin{eqnarray}\label{current}
j^{\mu}(x)&=&\varepsilon^{\mu\nu\alpha\beta}(\bar q_1(x)i\gamma_{\nu}b(x))D_{\alpha}(\bar{b
}(x)\gamma_{\beta}q_2(x)),
\end{eqnarray}
where $q_i$ stands for light quarks. Performing the parity transformation to the current, it satisfies the condition $Pj^{\mu}(x)P^{-1}=j^{\mu}(x)$.

Mass property is expected to be helpful for understanding the configuration of bottomonium-like resonances $Z_{b}$. In the hadronic scale, it is difficult to get reliable theoretical estimate for the mass using the perturbative QCD. Therefore, we need some non-perturative methods to describe the non-perturative phenomena. QCDSR ~\cite{svz,reinders,overview2,overview3,NielsenPR} is powerful since it is based on the fundamental QCD Lagrangian. We notice that in the case of $X(3872)$ with quantum numbers $J^{PC}=1^{++}$, the authors study its mass with QCDSR using the configuration $D^{0}\bar{D^{*0}}-D^{*0}\bar{D^{0}}$~\cite{Nielsen1} and the configuration $[cq]_{S=0}[\bar{c}\bar{q}]_{S=1}+[cq]_{S=1}[\bar{c}\bar{q}]_{S=0}$~\cite{Nielsen2}. Different from above situation, we consider the $Z_{b}(10610)$ with known quantum numbers $I^{G}J^{P}=1^{+}1^{+}$ with no definite charge conjugation. This work is devoted to investigate the masses of $Z_{b}(10610)$ and $Z_{b}(10650)$ in the QCDSR, both a $B\bar{B}^*$ molecular state and a $[bu][\bar{b}\bar{d}]$ tetraquark state being assumed.

The rest of the paper is organized as three parts. The QCDSR for the $Z_{b}(10610)$ is derived in Sec. \ref{sec2}, with contributions up to dimension six in the OPE. The numerical analysis is presented to extract the hadronic mass at the end of this section. Sec. \ref{sec3} is organized for the QCD sum rules of the $Z_{b}(10650)$, with numerical estimation of the mass. Sec. \ref{sec4} is the summary and conclusion.

\section{QCD sum rules for $Z_{b}(10610)$}\label{sec2}
\subsection{molecular state QCD sum rules}\label{sec2A}
In the previous work~\cite{Zhang}, $Z_{b}(10610)$ has been studied as a $B^{*}\bar{B}$ molecule with $J^{P}=1^{+}$ in the framework of QCDSR. However, the information of isospin and G-parity, which have been confirmed by the experiment, is omitted in constructing the interpolating current. In this subsection, the $Z_{b}(10610)$ resonance is considered by constructing a proper interpolator as a $B^{*}\bar{B}$ molecule with all the known quantum numbers $I^{G}J^{P}=1^{+}1^{+}$~\cite{10620}:
\begin{eqnarray}
j^{\mu}&=&\frac{1}{\sqrt{2}}[(\bar{u}i\gamma^{5}b)(\bar{b}\gamma_{\mu}d)+(\bar{u}_{a}\gamma_{\mu}b_{a})(\bar{b}_{b}i\gamma^{5}d_{b})].
\end{eqnarray}

In the QCDSR approach, the mass of the particle can be determined by considering the two-point correlation function
\begin{eqnarray}
\Pi^{\mu\nu}(q^{2})=i\int
d^{4}x\mbox{e}^{iq.x}\langle0|T[j^{\mu}(x)j^{\nu+}(0)]|0\rangle.
\end{eqnarray}
Lorentz covariance implies that the two-point correlation function can be generally parameterized as
\begin{eqnarray}
\Pi^{\mu\nu}(q^{2})=(\frac{q^{\mu}q^{\nu}}{q^{2}}-g^{\mu\nu})\Pi^{(1)}(q^{2})+\frac{q^{\mu}q^{\nu}}{q^{2}}\Pi^{(0)}(q^{2}).
\end{eqnarray}
The term proportional to $g_{\mu\nu}$ will be chosen to extract the
mass sum rule, since it
gets contributions only from the $1^{+}$ state.
The QCD sum rule attempts to link the hadron phenomenology with the
interactions of quarks and gluons. It contains three main
ingredients: an approximate description of the correlation function in terms
of intermediate states through the dispersion relation, a
description of the same correlation function in terms of QCD degrees of
freedom via an OPE, and a procedure for matching these two
descriptions and extracting the parameters that characterize the
hadronic state of interest.

In the phenomenological side, the correlation function is calculated by inserting a complete set of intermediate states. Parameterizing the coupling of the $Z$ state to the current $j^{\mu}$ as
\begin{eqnarray}
\langle 0|j_{\mu}|Z\rangle&=&\lambda \epsilon_{\mu}.
\end{eqnarray}
$\Pi^{(1)}(q^{2})$ can be expressed as
\begin{eqnarray}\label{ph}
\Pi^{(1)}(q^{2})=\frac{\lambda^{2}}{M_{Z}^{2}-q^{2}}+\frac{1}{\pi}\int_{s_{0}}
^{\infty}ds\frac{\mbox{Im}\Pi^{(1)\mbox{phen}}(s)}{s-q^{2}},
\end{eqnarray}
where $M_{Z}$ denotes the mass of the molecular state, and $s_0$ is the threshold parameter.

In the OPE side, $\Pi^{(1)}(q^{2})$ can be written as
\begin{eqnarray}\label{ope}
\Pi^{(1)}(q^{2})=\int_{4m_{b}^{2}}^{\infty}ds\frac{\rho^{OPE}(s)}{s-q^{2}},
\end{eqnarray}
where the spectral density is $\rho^{OPE}(s)=\frac{1}{\pi}\mbox{Im}\Pi^{\mbox{(1)}}(s)$.
Making quark-hadron duality assumption and a Borel transformation, the sum rule is obtained by matching the two sides:
\begin{eqnarray}\label{sr}
\lambda^{2}e^{-M_{Z}^{2}/M^{2}}&=&\int_{4m_{b}^{2}}^{s_{0}}ds\rho^{OPE}(s)e^{-s/M^{2}},
\end{eqnarray}
with $M^2$ the Borel parameter.

In the OPE side, we work at the leading order in $\alpha_{s}$
and consider vacuum condensates up to dimension six, with the similar
techniques in Refs.~\cite{technique}. In order to consider the isospin violation, we keep the terms which are linear in the light-quark
masses $m_{u}$ and $m_{d}$. After some tedious OPE calculations, the concrete forms of spectral densities can be derived:
\begin{eqnarray}
\rho^{OPE}(s)=\rho^{\mbox{pert}}(s)+\rho^{\langle\bar{q}q\rangle}(s)+\rho^{\langle
g^{2}G^{2}\rangle}(s)+\rho^{\langle
g\bar{q}\sigma\cdot G q\rangle}(s)+\rho^{\langle\bar{q}q\rangle^{2}}(s)+\rho^{\langle g^{3}G^{3}\rangle}(s),
\end{eqnarray}
with
\begin{eqnarray}\label{spectralmole}
\rho^{\mbox{pert}}(s)&=&\frac{3}{2^{12}\pi^{6}}\int_{\alpha_{min}}^{\alpha_{max}}\frac{d\alpha}{\alpha^{3}}\int_{\beta_{min}}^{1-\alpha}\frac{d\beta}{\beta^{3}}(1-\alpha-\beta)(1+\alpha+\beta)r(m_{b},s)^{4}
\nonumber\\&&{}
+\frac{3m_{b}}{2^{11}\pi^{6}}\int_{\alpha_{min}}^{\alpha_{max}}\frac{d\alpha}{\alpha^{3}}\int_{\beta_{min}}^{1-\alpha}\frac{d\beta}{\beta^{3}}(\alpha+\beta-1)(m_{u}\alpha^2+m_{d}\beta^2
\nonumber\\&&{}+m_{u}\alpha\beta+m_{d}\alpha\beta+3m_{u}\alpha+3m_{d}\beta)r(m_{b},s)^{3}
,\nonumber\\
\rho^{\langle\bar{q}q\rangle}(s)&=&-\frac{3\langle\bar{q}q\rangle}{2^{8}\pi^{4}}m_{b}\int_{\alpha_{min}}^{\alpha_{max}}\frac{d\alpha}{\alpha^{2}}\int_{\beta_{min}}^{1-\alpha}\frac{d\beta}{\beta^{2}}(\alpha+\beta)(1+\alpha+\beta)r(m_{b},s)^{2}
\nonumber\\&&{}
+\frac{3\langle\bar{q}q\rangle}{2^{8}\pi^{4}}(m_{u}+m_{d})\int_{\alpha_{min}}^{\alpha_{max}}\frac{d\alpha}{\alpha(1-\alpha)}[m_{Q}^{2}-\alpha(1-\alpha)s]^2
\nonumber\\&&{}
+\frac{3\langle\bar{q}q\rangle}{2^{6}\pi^{4}}m_{b}^2(m_{u}+m_{d})\int_{\alpha_{min}}^{\alpha_{max}}\frac{d\alpha}{\alpha}\int_{\beta_{min}}^{1-\alpha}\frac{d\beta}{\beta}r(m_{b},s)
\nonumber\\&&{}
-\frac{3\langle\bar{q}q\rangle}{2^{8}\pi^{4}}(m_{u}+m_{d})\int_{\alpha_{min}}^{\alpha_{max}}\frac{d\alpha}{\alpha}\int_{\beta_{min}}^{1-\alpha}\frac{d\beta}{\beta}r(m_{b},s)^2
,\nonumber\\
\rho^{\langle g^{2}G^{2}\rangle}(s)&=&\frac{\langle
g^{2}G^{2}\rangle}{2^{11}\pi^{6}}m_{b}^{2}\int_{\alpha_{min}}^{\alpha_{max}}d\alpha\int_{\beta_{min}}^{1-\alpha}\frac{d\beta}{\beta^{3}}(1-\alpha-\beta)(1+\alpha+\beta)r(m_{b},s)\nonumber\\&&{}
+\frac{\langle
g^{2}G^{2}\rangle}{2^{11}\pi^{6}}\int_{\alpha_{min}}^{\alpha_{max}}\frac{d\alpha}{\alpha}\int_{\beta_{min}}^{1-\alpha}\frac{d\beta}{\beta^{2}}(2\alpha+2\beta-1)r(m_{b},s)^{2}
,\nonumber\\
\rho^{\langle g\bar{q}\sigma\cdot G q\rangle}(s)&=&-\frac{3\langle
g\bar{q}\sigma\cdot G
q\rangle}{2^{7}\pi^{4}}m_{b}\int_{\alpha_{min}}^{\alpha_{max}}\frac{d\alpha}{\alpha}[m_{b}^{2}-\alpha(1-\alpha)s]
\nonumber\\&&{}
+\frac{3\langle g\bar{q}\sigma\cdot
Gq\rangle}{2^{8}\pi^{4}}m_{b}\int_{\alpha_{min}}^{\alpha_{max}}d\alpha\int_{\beta_{min}}^{1-\alpha}\frac{d\beta}{\beta}r(m_{b},s)
,\nonumber\\&&{}
+\frac{3\langle g\bar{q}\sigma\cdot
Gq\rangle}{2^{7}\pi^{4}}m_{b}\int_{\alpha_{min}}^{\alpha_{max}}d\alpha\int_{\beta_{min}}^{1-\alpha}\frac{d\beta}{\beta^2}(\alpha+\beta)r(m_{b},s)
,\nonumber\\
\rho^{\langle\bar{q}q\rangle^{2}}(s)&=&\frac{\langle\bar{q}q\rangle^{2}}{2^{4}\pi^{2}}m_{b}^{2}\sqrt{1-4m_{b}^{2}/s}
,\nonumber\\
\rho^{\langle g^{3}G^{3}\rangle}(s)&=&\frac{\langle
g^{3}G^{3}\rangle}{2^{12}\pi^{6}}m_{b}^{2}\int_{\alpha_{min}}^{\alpha_{max}}d\alpha\int_{\beta_{min}}^{1-\alpha}\frac{d\beta}{\beta^{3}}\alpha(1-\alpha-\beta)(1+\alpha+\beta)\nonumber\\&&{}
+\frac{\langle
g^{3}G^{3}\rangle}{2^{13}\pi^{6}}\int_{\alpha_{min}}^{\alpha_{max}}d\alpha\int_{\beta_{min}}^{1-\alpha}\frac{d\beta}{\beta^{3}}(1-\alpha-\beta)(1+\alpha+\beta)r(m_{b},s)
,
\end{eqnarray}
with $r(m_{b},s)=(\alpha+\beta)m_{b}^2-\alpha\beta s$. The integration limits are given by $\alpha_{min}=\Big(1-\sqrt{1-4m_{b}^{2}/s}\Big)/2$, $\alpha_{max}=\Big(1+\sqrt{1-4m_{b}^{2}/s}\Big)/2$, and $\beta_{min}=\alpha m_{b}^{2}/(s\alpha-m_{b}^{2})$.

Taking the derivative of Eq.(\ref{sr}) with respect to $\frac{1}{M^2}$ and then dividing by itself, we arrive at the mass of the molecular state
\begin{eqnarray}\label{sumrule}
M_{Z}^{2}&=&\int_{4m_{b}^{2}}^{s_{0}}ds\rho^{OPE}s
e^{-s/M^{2}}/
\int_{4m_{b}^{2}}^{s_{0}}ds\rho^{OPE}e^{-s/M^{2}}.
\end{eqnarray}

Before the numerical analysis of the equation (\ref{sumrule}), we first specify the input parameters. The quark masses are taken as $m_{u}=2.3~\mbox{MeV}$, $m_{d}=6.4~\mbox{MeV}$, and $m_{b}=(4.24\pm0.06)~\mbox{GeV}$ \cite{PDG}. The condensates are $\langle\bar{u}u\rangle=\langle\bar{d}d\rangle=\langle\bar{q}q\rangle=-(0.23\pm0.03)^{3}~\mbox{GeV}^{3}$,
$\langle g\bar{q}\sigma\cdot G
q\rangle=m_{0}^{2}~\langle\bar{q}q\rangle$,
$m_{0}^{2}=0.8~\mbox{GeV}^{2}$, $\langle
g^{2}G^{2}\rangle=0.88~\mbox{GeV}^{4}$, and $\langle
g^{3}G^{3}\rangle=0.045~\mbox{GeV}^{6}$~\cite{overview2}.
Complying with the standard procedure of the sum rule, the threshold $s_{0}$ and Borel parameter $M^{2}$ are varied to find the optimal stability window. There are two criteria (pole dominance and convergence of the OPE) for choosing the Borel parameter $M^{2}$ and threshold $s_{0}$.

\begin{figure}
\begin{center}
\begin{minipage}[c]{0.5\textwidth}
\centering
\includegraphics[width=\textwidth]{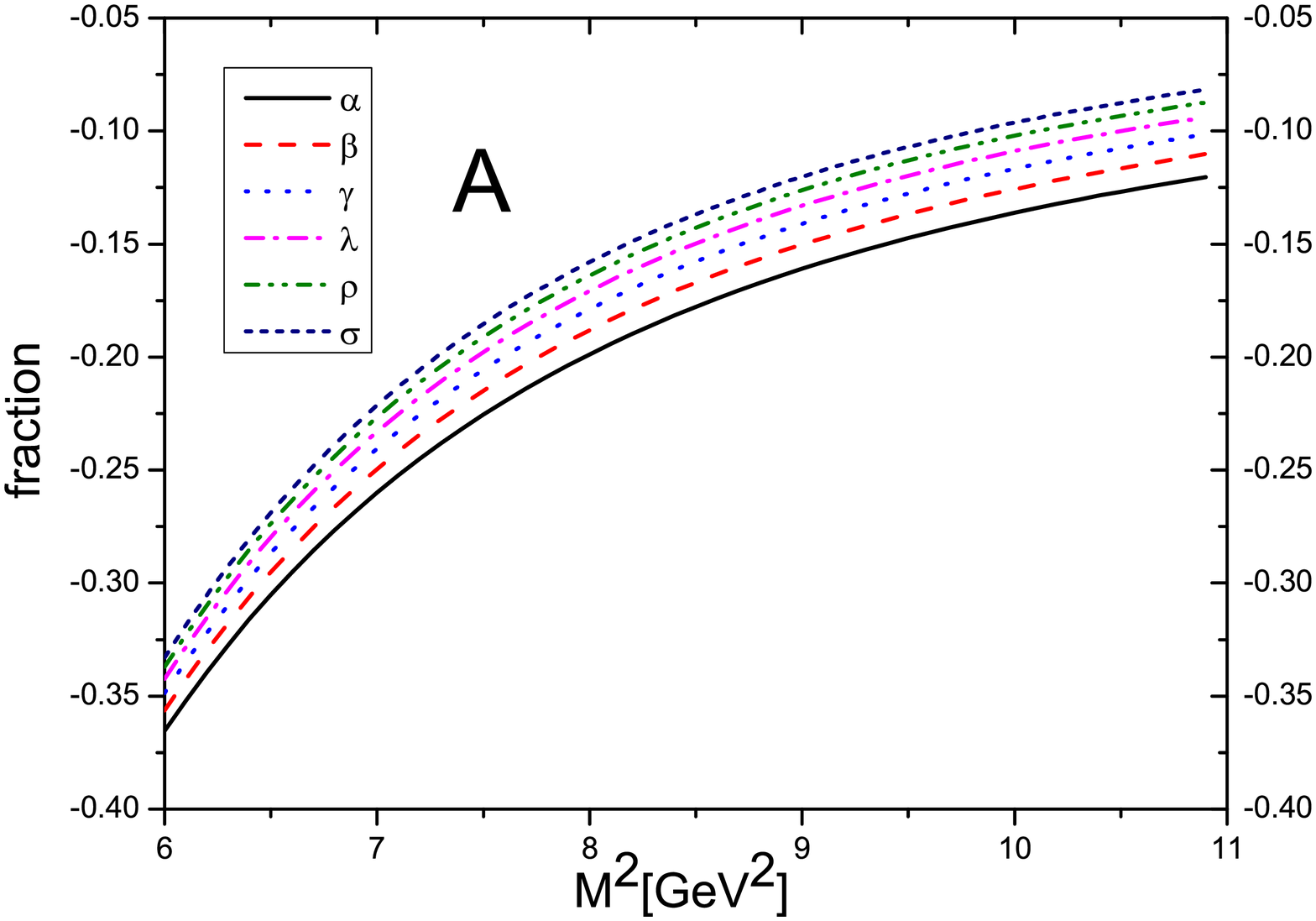}\\
(a)
\end{minipage}
\hspace{-0.1\textwidth}
\begin{minipage}[c]{0.5\textwidth}
\centering
\includegraphics[width=\textwidth]{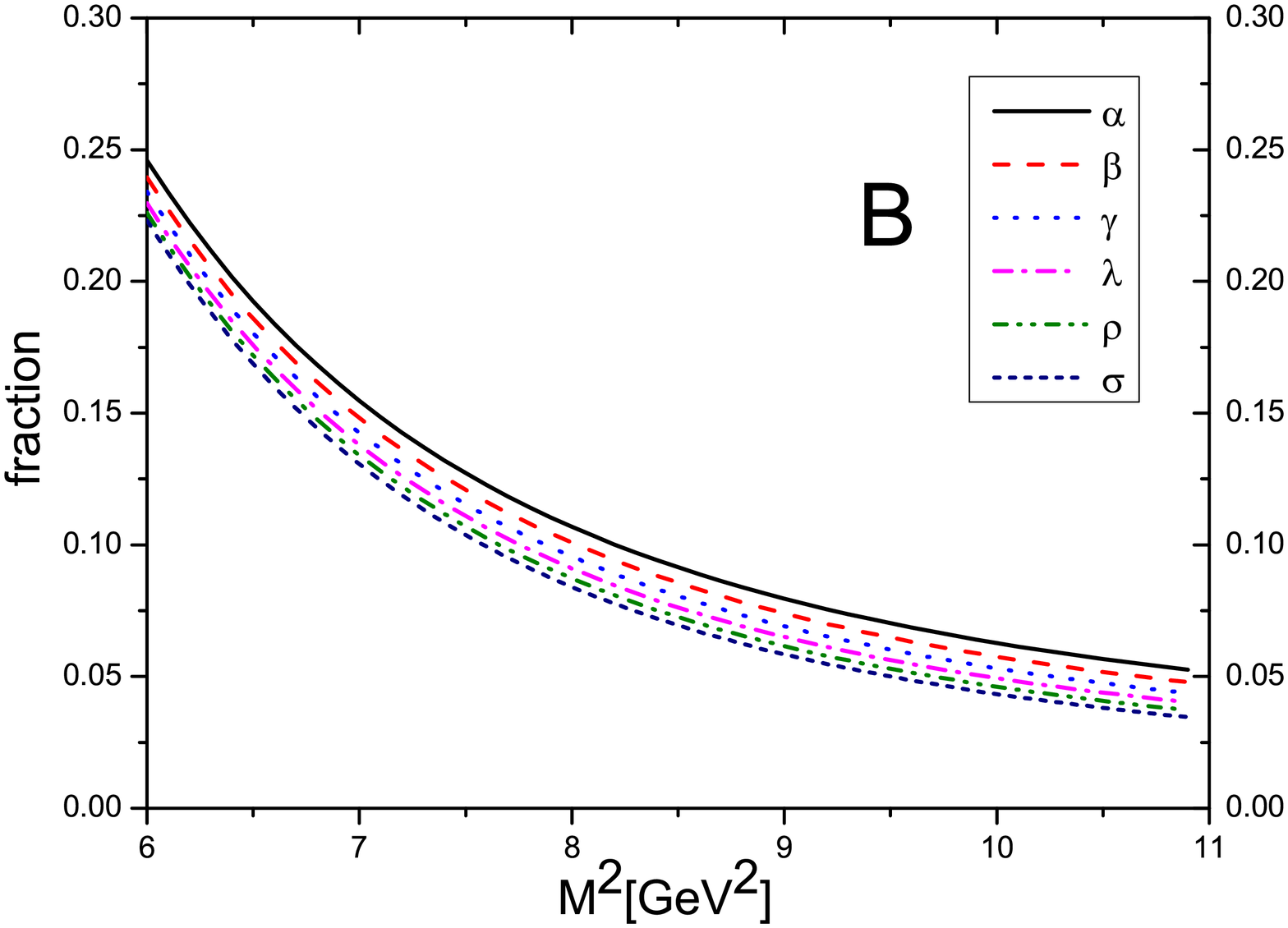}\\
(b)
\end{minipage}
\end{center}
\caption{The OPE convergence for the $I^{G}J^{P}=1^{+}1^{+}$ molecular state. The contributions from different terms with variation of the Borel parameter $M^2$ in the OPE. The $A$ and $B$  correspond to the contributions from the $D=5$ term and the $D=6$ term, respectively.  The notations $\alpha$, $\beta$, $\gamma$, $\lambda$, $\rho$ and $\sigma$ correspond to the threshold parameters $s_0=121.00\,\rm{GeV}^2$, $123.21\,\rm{GeV}^2$, $125.44\,\rm{GeV}^2$, $127.69\,\rm{GeV}^2$, $129.96\,\rm{GeV}^2$ and $132.25\,\rm{GeV}^2$, respectively.}\label{fig1}
\end{figure}

\begin{figure}
\centerline{\epsfysize=6.0truecm\epsfbox{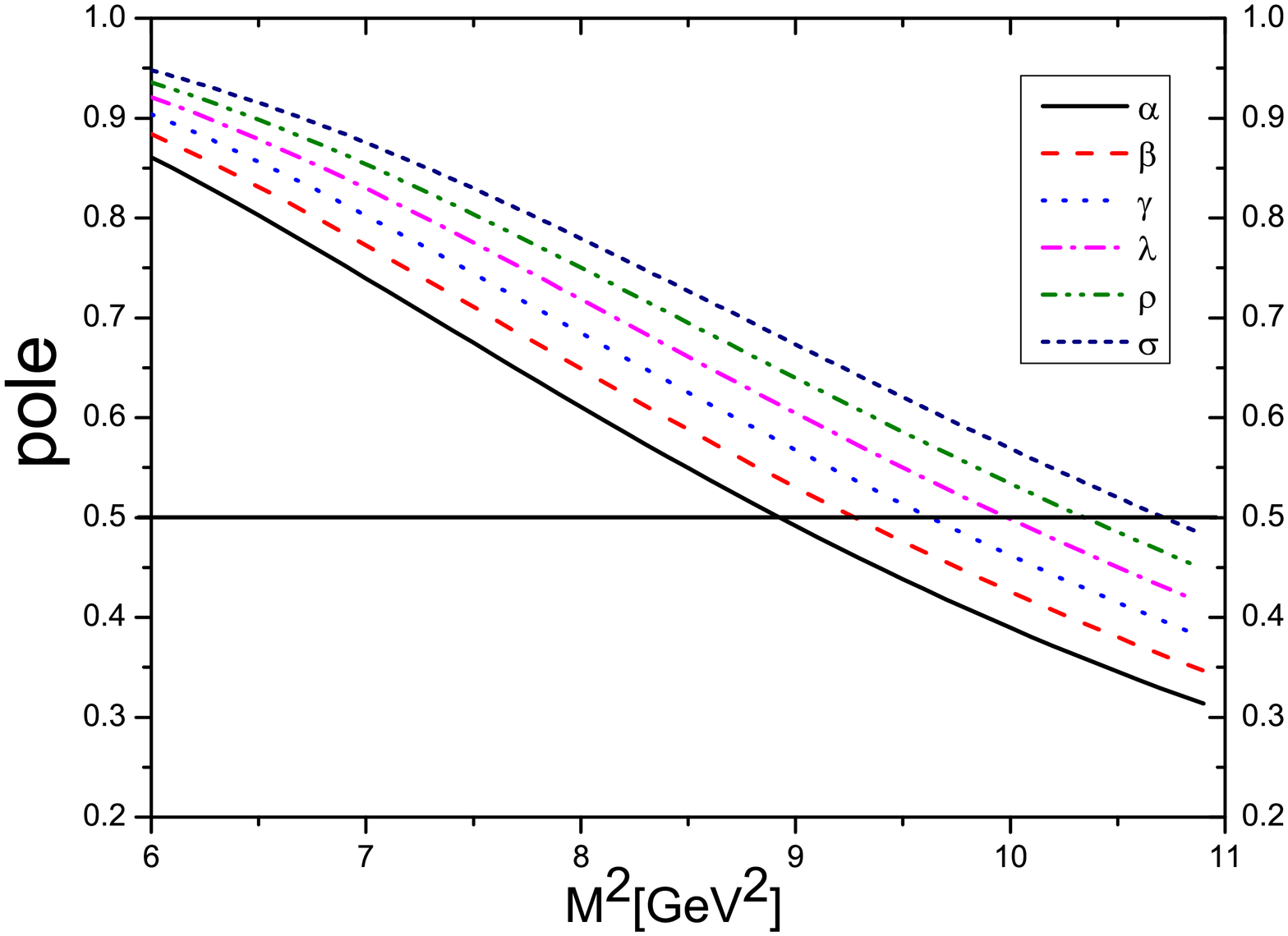}}
\caption{The contributions from the pole terms with variation of the Borel parameter $M^2$ in the case of molecular state. The notations $\alpha$, $\beta$, $\gamma$, $\lambda$, $\rho$ and $\sigma$ correspond to the threshold parameters $s_0=121.00\,\rm{GeV}^2$, $123.21\,\rm{GeV}^2$, $125.44\,\rm{GeV}^2$, $127.69\,\rm{GeV}^2$, $129.96\,\rm{GeV}^2$ and $132.25\,\rm{GeV}^2$, respectively.}\label{fig2}
\end{figure}

\begin{figure}
\centerline{\epsfysize=6.0truecm
\epsfbox{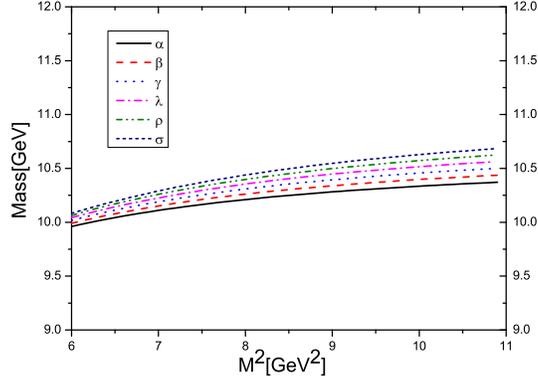}}\caption{
The mass of the $I^{G}J^{P}=1^{+}1^{+}$ molecular state as
a function of $M^2$ from sum rule (\ref{sumrule}). The notations $\alpha$, $\beta$, $\gamma$, $\lambda$, $\rho$ and $\sigma$ correspond to the threshold parameters $s_0=121.00\,\rm{GeV}^2$, $123.21\,\rm{GeV}^2$, $125.44\,\rm{GeV}^2$, $127.69\,\rm{GeV}^2$, $129.96\,\rm{GeV}^2$ and $132.25\,\rm{GeV}^2$, respectively.}\label{fig3}
\end{figure}

The contributions from the high dimension vacuum condensates in the OPE are shown in Fig.\ref{fig1}. We have used $s_0\geq 121\,\mbox{GeV}^2$. From this figure it can be seen that for $M^2\geq 8.10\,\mbox{GeV}^2$, the contribution of the dimension-$6$ condensate is less than $5\%$ of the total contribution and  the contribution of the dimension-$5$ condensate is less than $20\%$ of the total contribution, which indicate a good Borel convergence. Therefore, we fix the uniform lower value of $M^2$ in the sum rule window as $M^2_{min}= 8.10\,\mbox{GeV}^2$. The upper limit of $M^2$ is determined by imposing that the pole contribution should be larger than continuum contribution. Fig.\ref{fig2} shows that the contributions from the pole terms with variation of the Borel parameter $M^2$. We show in Table~\ref{tab:1} the values of $M^2_{max}$ for several values of $\sqrt{s_0}$. In Fig.\ref{fig3}, we show the molecular state mass, for different values of $\sqrt{s_0}$, in the relevant sum rule window. It can be seen that the mass is stable in the Borel window with the corresponding threshold $\sqrt{s_0}$. The final estimate of the $I^{G}J^{P}=1^{+}1^{+}$ molecular state is obtained as
\begin{eqnarray}
M_{Z} = (10.44\pm 0.23)~\mbox{GeV}.
\label{Zmass}
\end{eqnarray}

\begin{table}
\caption{Upper limits in the Borel window for the $I^{G}J^{P}=1^{+}1^{+}$ $B^*\bar{B}^*$ current obtained from the sum rule for different values of $\sqrt{s_0}$.}\label{tab:1}
\begin{center}
\begin{tabular}{|c|c|}
\hline
$\sqrt{s_0}~(\mbox{GeV})$ & $M^2_{max}(\mbox{GeV}^2)$\\ \hline
11.0 &8.9 \\ \hline
11.1 &9.2 \\ \hline
11.2 &9.6 \\ \hline
11.3 &9.9 \\ \hline
11.4 &10.3\\ \hline
11.5 &10.7\\ \hline
\end{tabular}
\end{center}
\end{table}

\subsection{tetraquark state QCD sum rules}
As pointed out in ref.~\cite{Guo}, if the charmonium-like tetraquark $X(3872)$ and $Z^+(4430)$ are really tetraquark states, the bottomonium-like tetraquark should exist. A possible current interpolating a $I^{G}J^{P}=1^{+}1^{+}$ tetraquark state with $[bd]_{S=0}$ and $[\bar{b}\bar{u}]_{S=1}$ fields is given by
\begin{eqnarray}
j_{\mu}={\epsilon_{abc}\epsilon_{dec}\over\sqrt{2}}[(d_a^T C\gamma_5
b_b) (\bar{u}_d\gamma_\mu C\bar{b}_e^T)-(d_a^T C \gamma_\mu b_b) (\bar{u}_d \gamma_5 C \bar{b}_e^T)]\;, \label{field2}
\end{eqnarray}
The spectral densities of the sum rule are obtained with the same standard procedure of the approach as what was done in subsection~\ref{sec2A} for the molecular current:
\begin{eqnarray}
\rho^{OPE}(s)=\rho^{\mbox{pert}}(s)+\rho^{\langle\bar{q}q\rangle}(s)+\rho^{\langle
g^{2}G^{2}\rangle}(s)+\rho^{\langle
g\bar{q}\sigma\cdot G q\rangle}(s)+\rho^{\langle\bar{q}q\rangle^{2}}(s)+\rho^{\langle g^{3}G^{3}\rangle}(s),
\end{eqnarray}
with
\begin{eqnarray}\label{spectraltetra}
\rho^{\mbox{pert}}(s)&=&\frac{1}{2^{10}\pi^{6}}\int_{\alpha_{min}}^{\alpha_{max}}\frac{d\alpha}{\alpha^{3}}\int_{\beta_{min}}^{1-\alpha}\frac{d\beta}{\beta^{3}}(1-\alpha-\beta)(1+\alpha+\beta)r(m_{b},s)^{4}
\nonumber\\&&{}
+\frac{m_{b}}{2^{9}\pi^{6}}\int_{\alpha_{min}}^{\alpha_{max}}\frac{d\alpha}{\alpha^{3}}\int_{\beta_{min}}^{1-\alpha}\frac{d\beta}{\beta^{3}}(\alpha+\beta-1)(m_{u}\alpha^2+m_{d}\beta^2
\nonumber\\&&{}+m_{u}\alpha\beta+m_{d}\alpha\beta+3m_{u}\alpha+3m_{d}\beta)r(m_{b},s)^{3}
,\nonumber\\
\rho^{\langle\bar{q}q\rangle}(s)&=&\frac{\langle\bar{q}q\rangle}{2^{6}\pi^{4}}(m_{u}+m_{d})\int_{\alpha_{min}}^{\alpha_{max}}\frac{d\alpha}{\alpha(1-\alpha)}[m_{Q}^{2}-\alpha(1-\alpha)s]^2
\nonumber\\&&{}
-\frac{\langle\bar{q}q\rangle}{2^{6}\pi^{4}}m_{b}\int_{\alpha_{min}}^{\alpha_{max}}\frac{d\alpha}{\alpha^{2}}\int_{\beta_{min}}^{1-\alpha}\frac{d\beta}{\beta^{2}}(\alpha+\beta)(1+\alpha+\beta)r(m_{b},s)^{2}
\nonumber\\&&{}
+\frac{\langle\bar{q}q\rangle}{2^{4}\pi^{4}}m_{b}^2(m_{u}+m_{d})\int_{\alpha_{min}}^{\alpha_{max}}\frac{d\alpha}{\alpha}\int_{\beta_{min}}^{1-\alpha}\frac{d\beta}{\beta}r(m_{b},s)
\nonumber\\&&{}
-\frac{\langle\bar{q}q\rangle}{2^{6}\pi^{4}}(m_{u}+m_{d})\int_{\alpha_{min}}^{\alpha_{max}}\frac{d\alpha}{\alpha}\int_{\beta_{min}}^{1-\alpha}\frac{d\beta}{\beta}r(m_{b},s)^2
,\nonumber\\
\rho^{\langle g^{2}G^{2}\rangle}(s)&=&\frac{\langle
g^{2}G^{2}\rangle}{3*2^{9}\pi^{6}}m_{b}^{2}\int_{\alpha_{min}}^{\alpha_{max}}d\alpha\int_{\beta_{min}}^{1-\alpha}\frac{d\beta}{\beta^{3}}(1-\alpha-\beta)(1+\alpha+\beta)r(m_{b},s)\nonumber\\&&{}
+\frac{\langle
g^{2}G^{2}\rangle}{3*2^{12}\pi^{6}}m_{b}^{2}\int_{\alpha_{min}}^{\alpha_{max}}\frac{d\alpha}{\alpha}\int_{\beta_{min}}^{1-\alpha}\frac{d\beta}{\beta^{2}}(1-\alpha-\beta)(3+\alpha+\beta)r(m_{b},s)\nonumber\\&&{}
+\frac{\langle
g^{2}G^{2}\rangle}{3*2^{10}\pi^{6}}\int_{\alpha_{min}}^{\alpha_{max}}\frac{d\alpha}{\alpha}\int_{\beta_{min}}^{1-\alpha}\frac{d\beta}{\beta^{2}}(2\alpha+2\beta-1)r(m_{b},s)^{2}
,\nonumber\\
\rho^{\langle g\bar{q}\sigma\cdot G q\rangle}(s)&=&-\frac{\langle
g\bar{q}\sigma\cdot G
q\rangle}{2^{5}\pi^{4}}m_{b}\int_{\alpha_{min}}^{\alpha_{max}}\frac{d\alpha}{\alpha}[m_{b}^{2}-\alpha(1-\alpha)s]
\nonumber\\&&{}
+\frac{\langle g\bar{q}\sigma\cdot
Gq\rangle}{2^{6}\pi^{4}}m_{b}\int_{\alpha_{min}}^{\alpha_{max}}\frac{d\alpha}{\alpha}\int_{\beta_{min}}^{1-\alpha}d\beta r(m_{b},s)
\nonumber\\&&{}
+\frac{\langle g\bar{q}\sigma\cdot
Gq\rangle}{2^{6}\pi^{4}}m_{b}\int_{\alpha_{min}}^{\alpha_{max}}d\alpha\int_{\beta_{min}}^{1-\alpha}\frac{d\beta}{\beta^2}(\alpha+\beta)r(m_{b},s)
\nonumber\\&&{}
-\frac{\langle g\bar{q}\sigma\cdot
Gq\rangle}{3*2^{8}\pi^{4}}m_{b}\int_{\alpha_{min}}^{\alpha_{max}}\frac{d\alpha}{\alpha^2}\int_{\beta_{min}}^{1-\alpha}d\beta(\alpha+\beta+1)r(m_{b},s)
,\nonumber\\
\rho^{\langle\bar{q}q\rangle^{2}}(s)&=&\frac{\langle\bar{q}q\rangle^{2}}{3*2^{2}\pi^{2}}m_{b}^{2}\sqrt{1-4m_{b}^{2}/s}
,\nonumber\\
\rho^{\langle g^{3}G^{3}\rangle}(s)&=&\frac{\langle
g^{3}G^{3}\rangle}{3*2^{10}\pi^{6}}m_{b}^{2}\int_{\alpha_{min}}^{\alpha_{max}}d\alpha\int_{\beta_{min}}^{1-\alpha}\frac{d\beta}{\beta^{3}}\alpha(1-\alpha-\beta)(1+\alpha+\beta)\nonumber\\&&{}
+\frac{\langle
g^{3}G^{3}\rangle}{3*2^{11}\pi^{6}}\int_{\alpha_{min}}^{\alpha_{max}}d\alpha\int_{\beta_{min}}^{1-\alpha}\frac{d\beta}{\beta^{3}}(1-\alpha-\beta)(1+\alpha+\beta)r(m_{b},s)
.
\end{eqnarray}

\begin{table}
\caption{Upper limits in the Borel window for the $I^{G}J^{P}=1^{+}1^{+}$ tetraquark current obtained from the sum rule for different values of $\sqrt{s_0}$.}\label{tab:2}
\begin{center}
\begin{tabular}{|c|c|}
\hline
$\sqrt{s_0}(\mbox{GeV})$&$M^2_{max}(\mbox{GeV}^2)$\\ \hline
11.0 &8.7 \\ \hline
11.1 &9.1 \\ \hline
11.2 &9.4 \\ \hline
11.3 &9.8 \\ \hline
11.4 &10.2\\ \hline
11.5 &10.6\\ \hline
\end{tabular}
\end{center}
\end{table}

\begin{figure}
\begin{center}
\begin{minipage}[c]{0.5\textwidth}
\centering
\includegraphics[width=\textwidth]{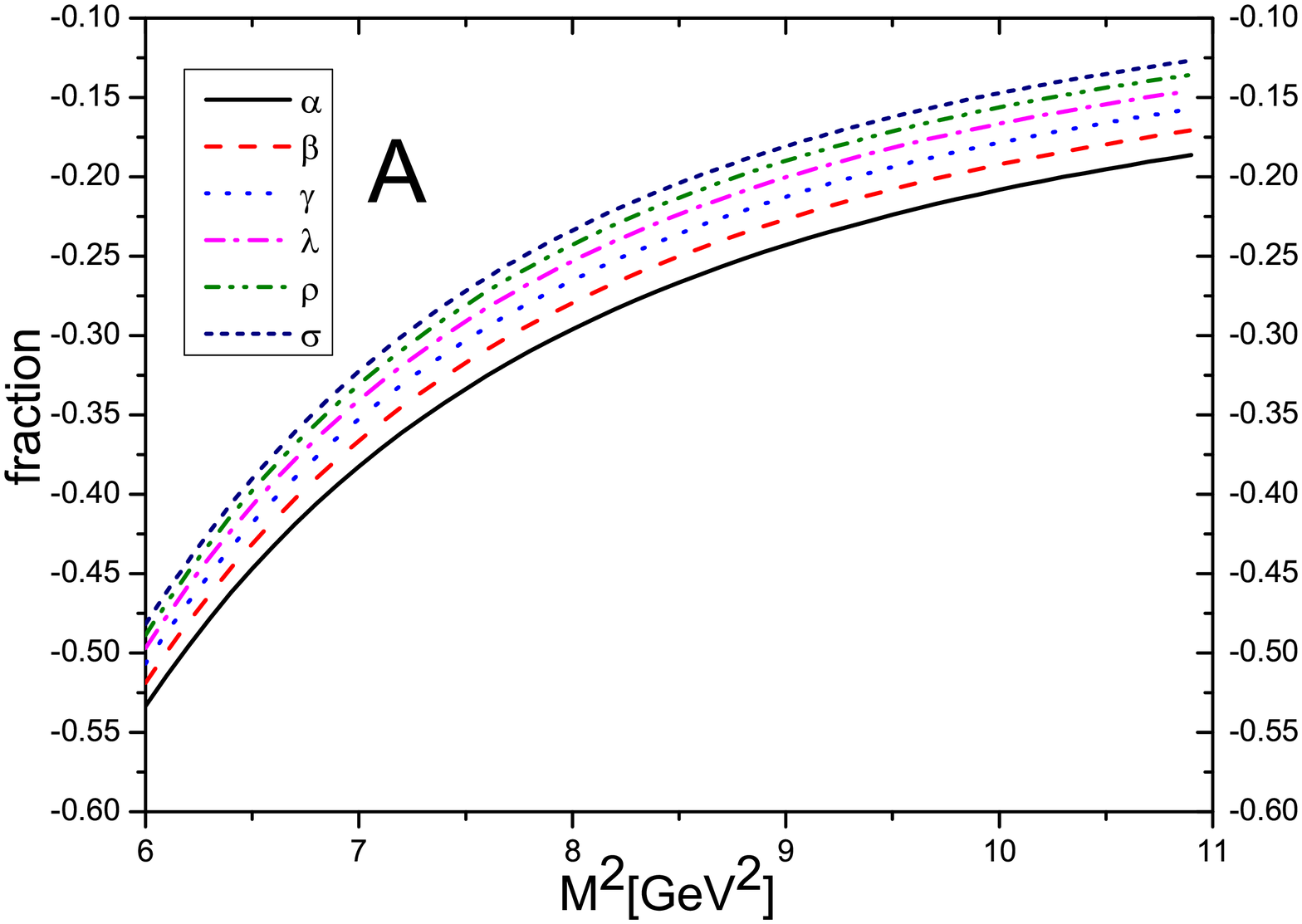}\\
(a)
\end{minipage}
\hspace{-0.1\textwidth}
\begin{minipage}[c]{0.5\textwidth}
\centering
\includegraphics[width=\textwidth]{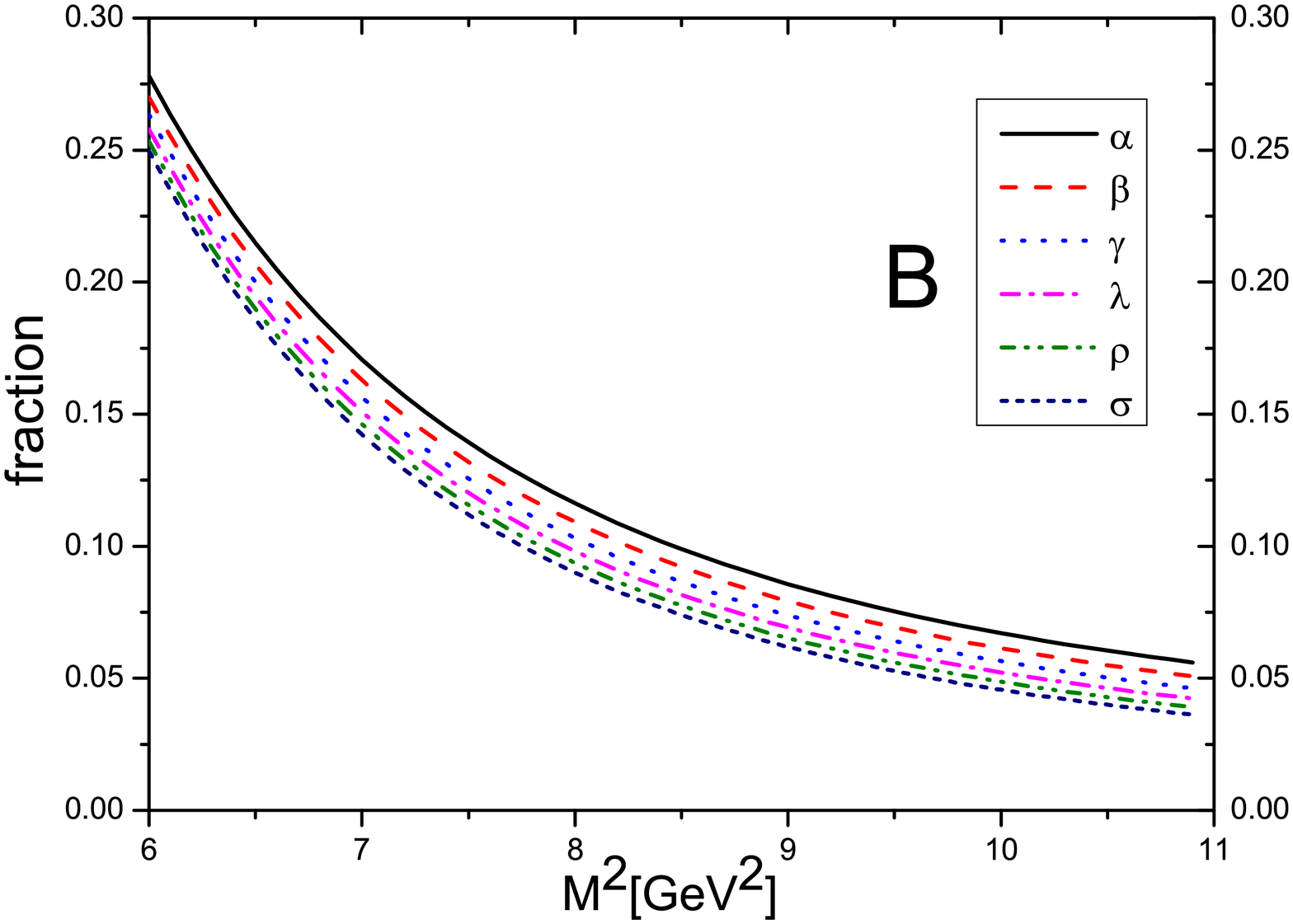}\\
(b)
\end{minipage}
\end{center}
\caption{The OPE convergence for the $I^{G}J^{P}=1^{+}1^{+}$ tetraquark state. The contributions from different terms with variation of the Borel parameter $M^2$ in the OPE. The $A$ and $B$  correspond to the contributions from the $D=5$ term and the $D=6$ term, respectively. The notations $\alpha$, $\beta$, $\gamma$, $\lambda$, $\rho$ and $\sigma$ correspond to the threshold parameters $s_0=121.00\,\rm{GeV}^2$, $123.21\,\rm{GeV}^2$, $125.44\,\rm{GeV}^2$, $127.69\,\rm{GeV}^2$, $129.96\,\rm{GeV}^2$ and $132.25\,\rm{GeV}^2$, respectively.}\label{fig4}
\end{figure}

\begin{figure}
\centerline{\epsfysize=6.0truecm\epsfbox{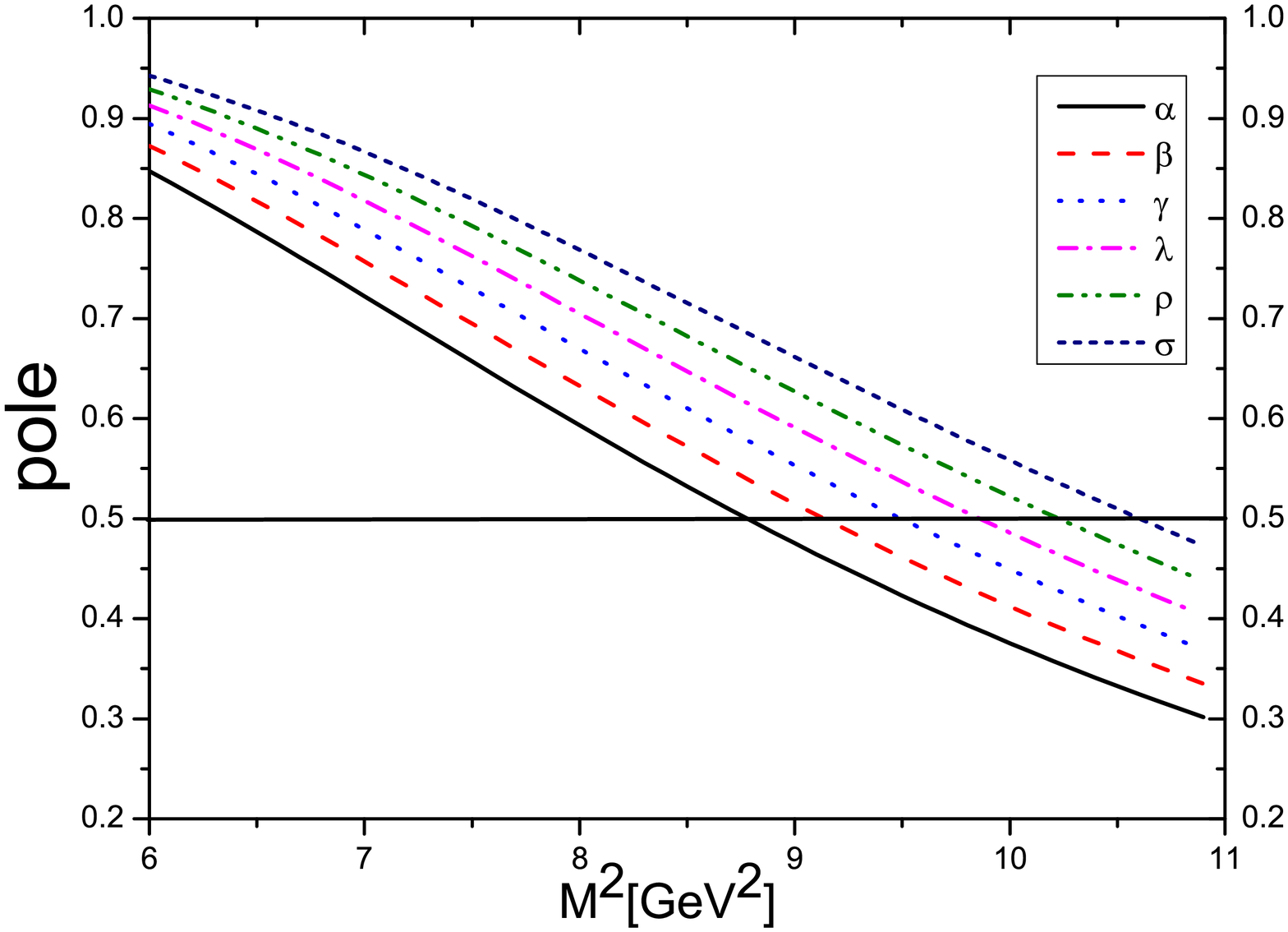}}
\caption{The contributions from the pole terms with variation of the Borel parameter $M^2$ in the case of tetraquark state. The notations $\alpha$, $\beta$, $\gamma$, $\lambda$, $\rho$ and $\sigma$ correspond to the threshold parameters $s_0=121.00\,\rm{GeV}^2$, $123.21\,\rm{GeV}^2$, $125.44\,\rm{GeV}^2$, $127.69\,\rm{GeV}^2$, $129.96\,\rm{GeV}^2$ and $132.25\,\rm{GeV}^2$, respectively.}\label{fig5}
\end{figure}

\begin{figure}
\centerline{\epsfysize=6.0truecm
\epsfbox{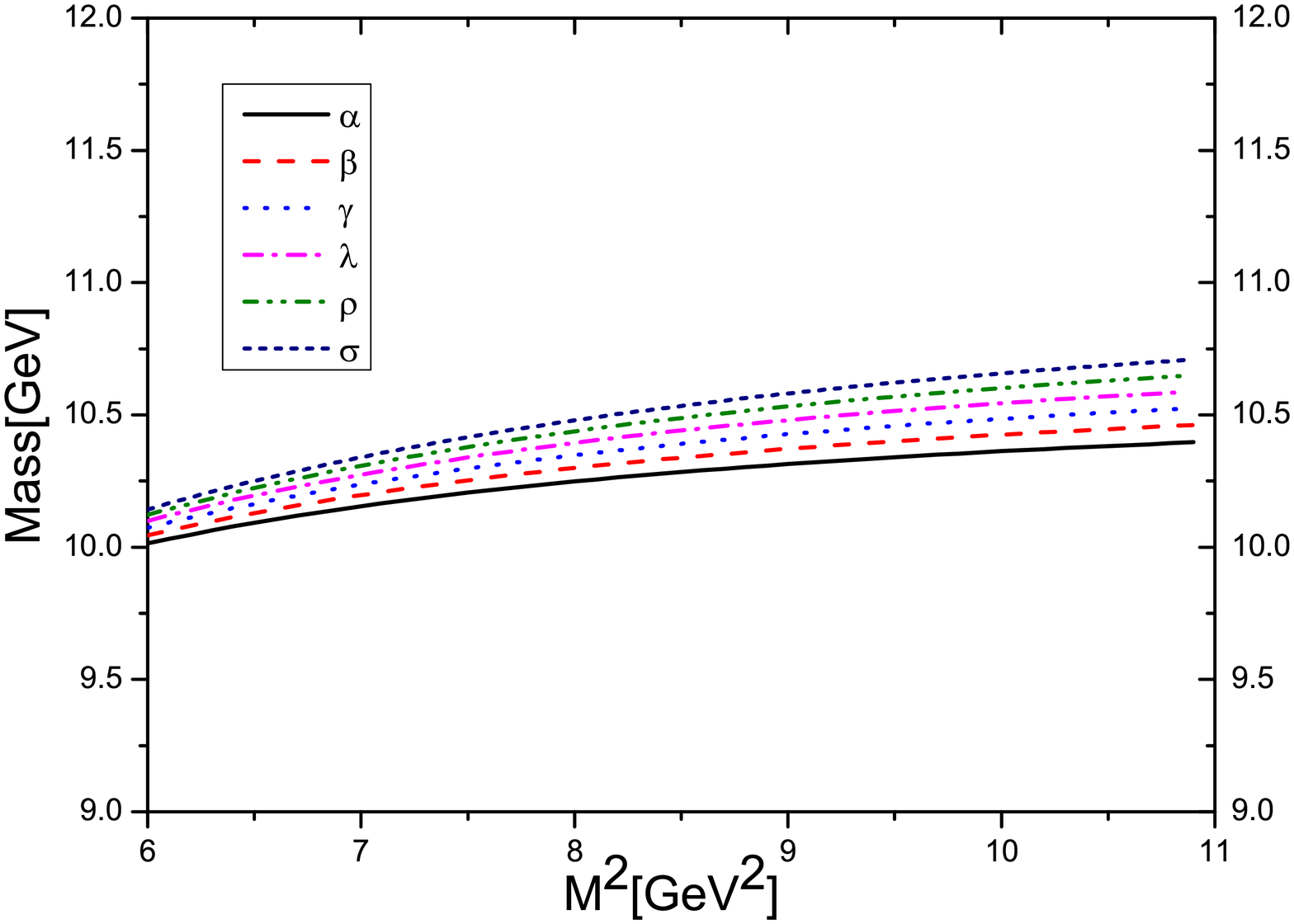}}\caption{
The mass of the tetraquark state as
a function of $M^2$. The notations $\alpha$, $\beta$, $\gamma$, $\lambda$, $\rho$ and $\sigma$ correspond to the threshold parameters $s_0=121.00\,\rm{GeV}^2$, $123.21\,\rm{GeV}^2$, $125.44\,\rm{GeV}^2$, $127.69\,\rm{GeV}^2$, $129.96\,\rm{GeV}^2$ and $132.25\,\rm{GeV}^2$, respectively.}\label{fig6}
\end{figure}
The mass sum rule is obtained as the same form of Eq.(\ref{sumrule}).
In this case, from Fig.\ref{fig4} we see that we obtain a reasonable OPE convergence for $M^2_{min}= 8.90\,\mbox{GeV}^2$. From this figure it can be seen that for $M^2\geq 8.90\,\mbox{GeV}^2$, the contribution of the dimension-$6$ condensate is less than $9\%$ of the total contribution and  the contribution of the dimension-$5$ condensate is less than $25\%$ of the total contribution. Fig.\ref{fig5} shows that the contributions from the pole terms with variation of the Borel parameter $M^2$. The upper limits of $M^2$ for each value of $\sqrt{s_0}$ are given in Table~\ref{tab:2}. The table also indicates that $\sqrt{s_{0}}\ge 11.1\,\mbox{GeV}$ to ensure $M^2_{max}\ge M^2_{min}$.  In Fig.\ref{fig6}, we show the mass $M_{Z}$ depending on the Borel mass for several threshold values $\sqrt{s_0}$.
It can be seen that we get a good Borel stability for $M_Z$.
The numerical result is
\begin{eqnarray}
M_{Z} = (10.5\pm 0.19)~\mbox{GeV}.
\label{Zmass1}
\end{eqnarray}

It is noticed that the result is very close to the one in Eq. (\ref{Zmass}). The reasons lies in the following arguments: The mass sum rule is expressed as a division of two parts as Eq.(\ref{sumrule}). However, in the two interpolating currents involved in the paper, the dominant contributions to the sum rules are the perturbative term and $D=3$ condensate term, which are proportional by $3/4$ that will be canceled in the process of division for both cases. The deviation comes from the $D=4$ and $D=5$ condensates terms which play subdominant roles in the result. It tacitly suggests that two-point sum rules is unable to give explicit distinction whether the exotic is a molecular state or a tetraquark state.

\subsection{General analysis on the mass sum rule for a four-quark state}

From the above study, we notice that for a four-quark state, there are two main candidates for its possible configuration which are a molecular state and a tetraquark state. In consideration of the requirement of the quantum numbers and the color singlet, the general form can be written as $\bar q_1\Gamma_i Q\bar Q\Gamma_jq_2$ for a molecular state and $\epsilon_{abc}\epsilon_{dec}(q_1^{aT} C\Gamma_iQ^b) (\bar q_2^d\Gamma_j C\bar Q^{eT})$ for a tetraquark. In the expressions $\Gamma_{i(j)}$ is a general form of a gamma matrix, and $\epsilon_{abc(def)}$ is the usual antisymmetric three order tensor with $a,b,c,d,e$ being color indices.

For a concrete form, there are ten possible combinations of $\Gamma_{i(j)}$ for quantum numbers $J^{P}=0^{-}$, $J^{P}=0^{+}$, $J^{P}=1^{-}$ and $J^{P}=1^{+}$. The ten possible currents for the two configurations are shown in Tab.\ref{tab:3}.
\begin{table}
\caption{Possible combinations of $\Gamma_{i(j)}$ for different $J^{P}$ quantum numbers. In each row, $\bar q_1\Gamma_i Q\bar Q\Gamma_jq_2$ stands for a molecular state and $\epsilon_{abc}\epsilon_{dec}(q_1^{aT} C\Gamma_iQ^b) (\bar q_2^d\Gamma_j C\bar Q^{eT})$ stands for a tetraquark state.}\label{tab:3}
\begin{center}
\begin{tabular}{|c|c|c|}
\hline
$J^{P}$ & $\gamma_{i}$ & $\gamma_{j}$\\ \hline
$J^{P}=0^{-}$&$\gamma_{5}$ & $1$\\ \hline
$J^{P}=0^{-}$&$\gamma_{\mu}$ &$\gamma_{\mu}\gamma_{5}$ \\ \hline
$J^{P}=0^{+}$&$\gamma_{5}$ &$\gamma_{5}$ \\ \hline
$J^{P}=0^{+}$&$1$ &$1$\\ \hline
$J^{P}=0^{+}$&$\gamma_{\mu}$ &$\gamma_{\mu}$ \\ \hline
$J^{P}=0^{+}$&$\gamma_{\mu}\gamma_{5}$ &$\gamma_{\mu}\gamma_{5}$\\ \hline
$J^{P}=1^{-}$&$1$ &$\gamma_{\mu}$\\ \hline
$J^{P}=1^{-}$&$\gamma_{5}$ & $\gamma_{\mu}\gamma_{5}$\\ \hline
$J^{P}=1^{+}$&$\gamma_{5}$ & $\gamma_{\mu}$\\ \hline
$J^{P}=1^{+}$&$1$ & $\gamma_{\mu}\gamma_{5}$\\ \hline
\end{tabular}
\end{center}
\end{table}
To get the sum rule, both the dispersion relationship and the quark-hadron duality approximation are used. The phenomenological part can be expressed as the integral of the spectral density:
\begin{eqnarray}\label{sr}
\lambda^{2}e^{-M_{Z}^{2}/M^{2}}&=&\int_{4m_{Q}^{2}}^{s_{0}}ds\rho^{OPE}(s)e^{-s/M^{2}},
\end{eqnarray}

After some tedious OPE calculations, the concrete forms of spectral densities can be derived:
\begin{eqnarray}
\rho^{OPE}(s)=\rho^{\mbox{pert}}(s)+\rho^{\langle\bar{q}q\rangle}(s)+\rho^{higherorder}(s).
\end{eqnarray}
To get the mass sum rule, we need to take the derivative of Eq.(\ref{sr}) with respect to $\frac{1}{M^2}$ and then divide
by itself. The final result is expressed as follows:
\begin{eqnarray}
M_{Z}^{2}&=&\int_{4m_{Q}^{2}}^{s_{0}}ds\rho^{OPE}s
e^{-s/M^{2}}/
\int_{4m_{Q}^{2}}^{s_{0}}ds\rho^{OPE}e^{-s/M^{2}}.
\end{eqnarray}

A simple calculation shows that the two dominant contributions $\rho^{\mbox{pert}}(s)$ and $\rho^{\langle\bar{q}q\rangle}(s)$ are same for the two configurations except for a factor. As in the analysis of the sum rules, Borel transformation is used to suppress both the higher resonance contribution and make the OPE convergence better. One of the criteria for the choice of the Borel parameter is that in the working region the higher order contributions of the vacuum condensates are required to be less than about $20\%$. However, the errors of the sum rule come from the variation of the Borel parameter, the threshold and the input parameter, which may bring an uncertainty bigger than $20\%$. Thus, deviation from second order vacuum condensate contributions can not give a definite judge on the fact which configurations can be in accordance with the exotic state concerned.

The two different configurations are quite different and are governed by different dynamics. Unfortunately, the mass sum rules cannot give us the useful detailed information on this deviation. Therefore, thorough understanding of the internal structure of a exotic state in the framework of QCDSR requires to investigate various decay processes that may contain more detailed information of the state.
\section{QCD sum rules for $Z_{b}(10650)$}\label{sec3}
\subsection{molecular state QCD sum rules}\label{sec3A}
Due to the mass of $Z_{b}(10650)$ lies close to the $B^*B^{*}$ threshold, $B^*\bar{B}^*$ molecule seems to be a natural candidate. Based on $B^{*}\bar{B}^{*}$ molecular type structure, $Z_{b}(10650)$ is studied within one-boson-exchange model and XEFT theory in Ref.~\cite{Voloshin,Sun,mehen}. In order to investigate the state of this configuration with QCDSR, we construct a possible interpolator:
\begin{eqnarray}\label{current1}
j^{\mu}(x)&=&\varepsilon^{\mu\nu\alpha\beta}(\bar u(x)i\gamma_{\nu}b(x))D_{\alpha}(\bar{b
}(x)\gamma_{\beta}d(x)).
\end{eqnarray}

Following the standard procedure of QCDSR, we obtain the concrete forms of spectral densities as follows:
\begin{eqnarray}
\rho^{OPE}(s)=\rho^{\mbox{pert}}(s)+\rho^{\langle\bar{q}q\rangle}(s)+\rho^{\langle
g^{2}G^{2}\rangle}(s)+\rho^{\langle
g\bar{q}\sigma\cdot G q\rangle}(s)+\rho^{\langle\bar{q}q\rangle^{2}}(s)+\rho^{\langle g^{3}G^{3}\rangle}(s),
\end{eqnarray}
with
\begin{eqnarray}\label{spectralmole2}
\rho^{\mbox{pert}}(s)&=&\frac{3}{5*2^{11}\pi^{6}}\int_{\alpha_{min}}^{\alpha_{max}}\frac{d\alpha}{\alpha^{3}}\int_{\beta_{min}}^{1-\alpha}\frac{d\beta}{\beta^{4}}({\alpha}\beta+\beta^2-2{\alpha}-\beta)r(m_{b},s)^{5}
,\nonumber\\
\rho^{\langle\bar{q}q\rangle}(s)&=&\frac{\langle\bar{q}q\rangle}{2^{6}\pi^{4}}m_{b}\int_{\alpha_{min}}^{\alpha_{max}}\frac{d\alpha}{\alpha(1-\alpha)^3}[m_{b}^{2}-\alpha(1-\alpha)s]^3
\nonumber\\&&{}+\frac{\langle\bar{q}q\rangle}{2^{7}\pi^{4}}m_{b}\int_{\alpha_{min}}^{\alpha_{max}}\frac{d\alpha}{\alpha^{2}}\int_{\beta_{min}}^{1-\alpha}\frac{d\beta}{\beta^{3}}({\alpha}\beta+\beta^2-2{\alpha}-2\beta)r(m_{b},s)^{3}
,\nonumber\\
\rho^{\langle g^{2}G^{2}\rangle}(s)&=&\frac{\langle
g^{2}G^{2}\rangle}{3*2^{11}\pi^{6}}\int_{\alpha_{min}}^{\alpha_{max}}\frac{d\alpha(2\alpha+1)}{\alpha(\alpha-1)^3}[m_{b}^{2}-\alpha(1-\alpha)s]^3
\nonumber\\&&{}+\frac{\langle
g^{2}G^{2}\rangle}{3*2^{12}\pi^{6}}m_{b}^{4}\int_{\alpha_{min}}^{\alpha_{max}}\frac{d\alpha}{\alpha^4}\int_{\beta_{min}}^{1-\alpha}\frac{d\beta}{\beta^3}(1-\alpha-\beta)^2(5{\alpha}^4+2{\alpha}^3\beta
\nonumber\\&&{}-2{\alpha}^3+7{\alpha}\beta^3+4\beta^4+8\beta^3)r(m_{b},s)
\nonumber\\&&{}-\frac{\langle
g^{2}G^{2}\rangle}{3*2^{12}\pi^{6}}m_{b}^{2}\int_{\alpha_{min}}^{\alpha_{max}}\frac{d\alpha}{\alpha^4}\int_{\beta_{min}}^{1-\alpha}\frac{d\beta}{\beta^4}({\alpha}^6+3{\alpha}^5\beta-6{\alpha}^5-18{\alpha}^4\beta
\nonumber\\&&{}+21{\alpha}^4+10{\alpha}^3\beta^3-9{\alpha}^3\beta^2+15{\alpha}^3\beta-10{\alpha}^3+12{\alpha}^2\beta^4)r(m_{b},s)^{2}
\nonumber\\&&{}+\frac{\langle
g^{2}G^{2}\rangle}{3*2^{12}\pi^{6}}\int_{\alpha_{min}}^{\alpha_{max}}\frac{d\alpha}{\alpha^3}\int_{\beta_{min}}^{1-\alpha}\frac{d\beta}{\beta^3}(2{\alpha}^3+17{\alpha}^2-4{\alpha}\beta)r(m_{b},s)^3
,\nonumber\\
\rho^{\langle g\bar{q}\sigma\cdot G q\rangle}(s)&=&\frac{3\langle
g\bar{q}\sigma\cdot G q\rangle}{2^{9}\pi^{4}}m_{b}\int_{\alpha_{min}}^{\alpha_{max}}d\alpha\frac{(7\alpha-5)}{\alpha(\alpha-1)^2}[m_{b}^{2}-\alpha(1-\alpha)s]^2
\nonumber\\&&{}+\frac{3\langle
g\bar{q}\sigma\cdot G q\rangle}{2^{7}\pi^{4}}m_{b}\int_{\alpha_{min}}^{\alpha_{max}}\frac{d\alpha}{(\alpha-1)}{\alpha}s[m_{b}^{2}-\alpha(1-\alpha)s]
\nonumber\\&&{}
-\frac{3\langle g\bar{s}\sigma\cdot
Gs\rangle}{2^{9}\pi^{4}}m_{b}\int_{\alpha_{min}}^{\alpha_{max}}\frac{d\alpha}{\alpha^2}\int_{\beta_{min}}^{1-\alpha}\frac{d\beta}{\beta^2}(\alpha^2+\beta^2-2\beta)r(m_{b},s)^2
,\nonumber\\
\rho^{\langle\bar{q}q\rangle^{2}}(s)&=&-\frac{\langle\bar{q}q\rangle^{2}}{3*2^{5}\pi^{2}}(8m_{b}^{4}+m_{b}^{2}s)\sqrt{1-4m_{b}^{2}/s}
,\nonumber\\
\rho^{\langle g^{3}G^{3}\rangle}(s)&=&-\frac{7\langle
g^{3}G^{3}\rangle}{3*2^{16}\pi^{6}}m_{b}^{2}\int_{\alpha_{min}}^{\alpha_{max}}d\alpha\int_{\beta_{min}}^{1-\alpha}\frac{d\beta}{\beta^{4}}(7\alpha^3+27\alpha^2\beta-39\alpha^2+21\alpha\beta^2
\nonumber\\&&{}-30{\alpha}\beta+9{\alpha}+\beta^3-3\beta^2+3\beta-1)r(m_{b},s)
\nonumber\\&&{}+\frac{\langle
g^{3}G^{3}\rangle}{2^{13}\pi^{6}}\int_{\alpha_{min}}^{\alpha_{max}}\frac{d\alpha}{\alpha}\int_{\beta_{min}}^{1-\alpha}\frac{d\beta}{\beta^{4}}(1-\alpha-\beta)^2r(m_{b},s)^2
.
\end{eqnarray}

\begin{figure}
\begin{center}
\begin{minipage}[c]{0.5\textwidth}
\centering
\includegraphics[width=\textwidth]{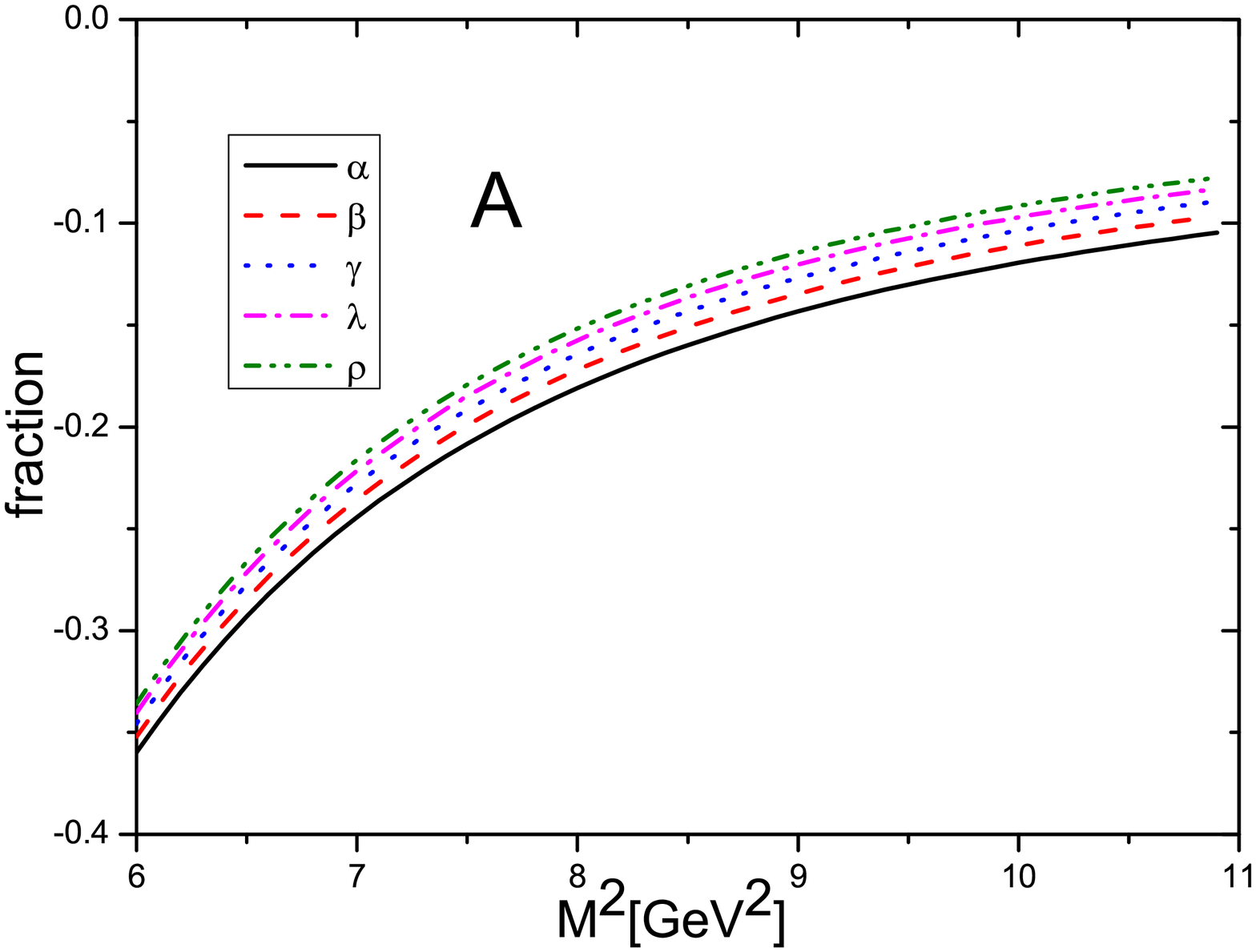}\\
(a)
\end{minipage}
\hspace{-0.1\textwidth}
\begin{minipage}[c]{0.5\textwidth}
\centering
\includegraphics[width=\textwidth]{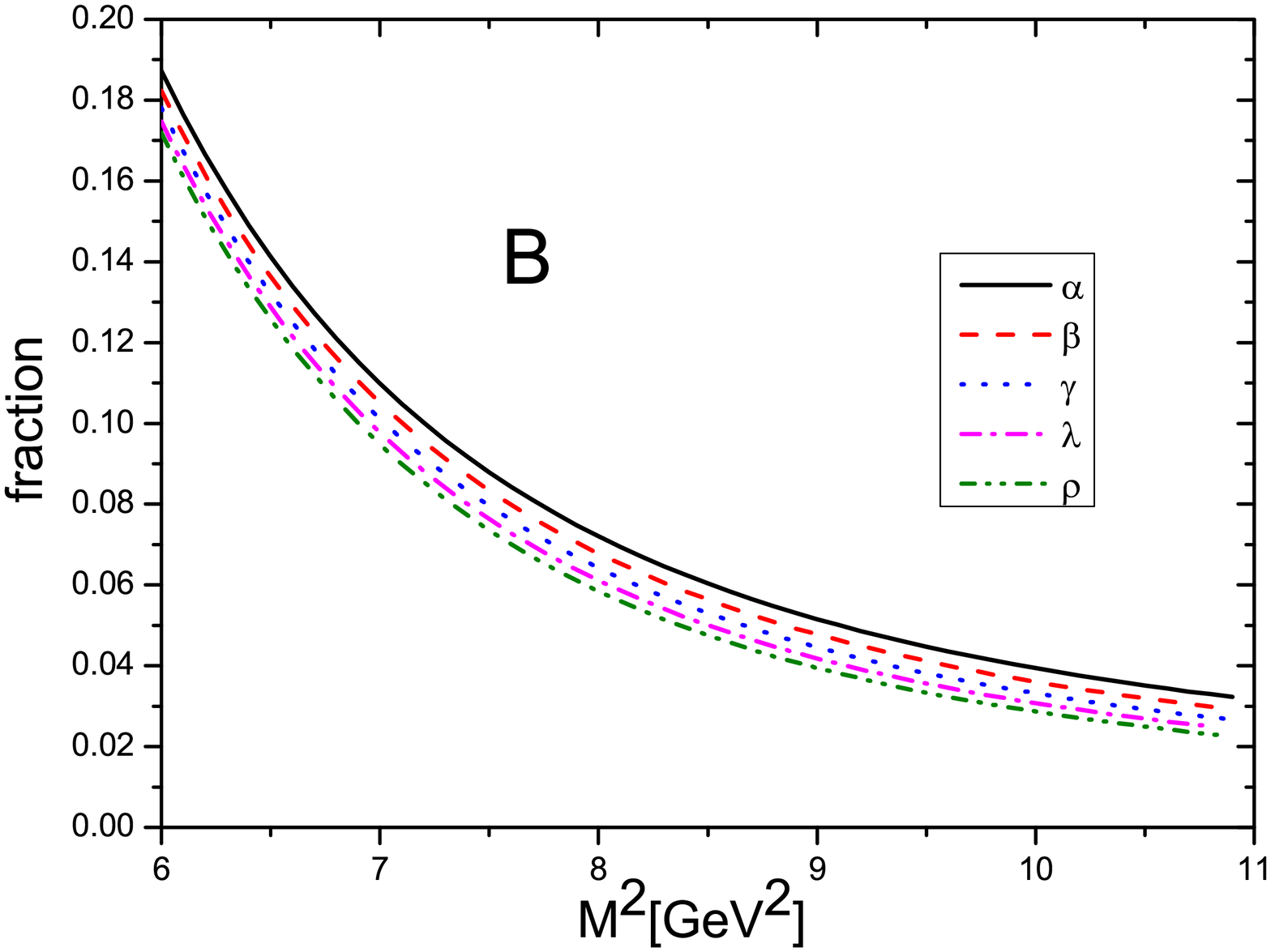}\\
(b)
\end{minipage}
\end{center}
\caption{The OPE convergence for the axial-vector $B^{*}\bar{B^{*}}$ molecular states. The contributions from different terms with variation of the Borel parameter $M^2$ in the OPE. The $A$ and $B$ correspond to the contributions from the $D=5$ term and the $D=6$ term, respectively. The notations $\alpha$, $\beta$, $\gamma$, $\lambda$ and $\rho$ correspond to the threshold parameters $s_0=121.00\,\rm{GeV}^2$, $123.21\,\rm{GeV}^2$, $125.44\,\rm{GeV}^2$, $127.69\,\rm{GeV}^2$ and $129.96\,\rm{GeV}^2$, respectively.}\label{fig7}
\end{figure}

\begin{figure}
\centering
\includegraphics[totalheight=5cm,width=7cm]{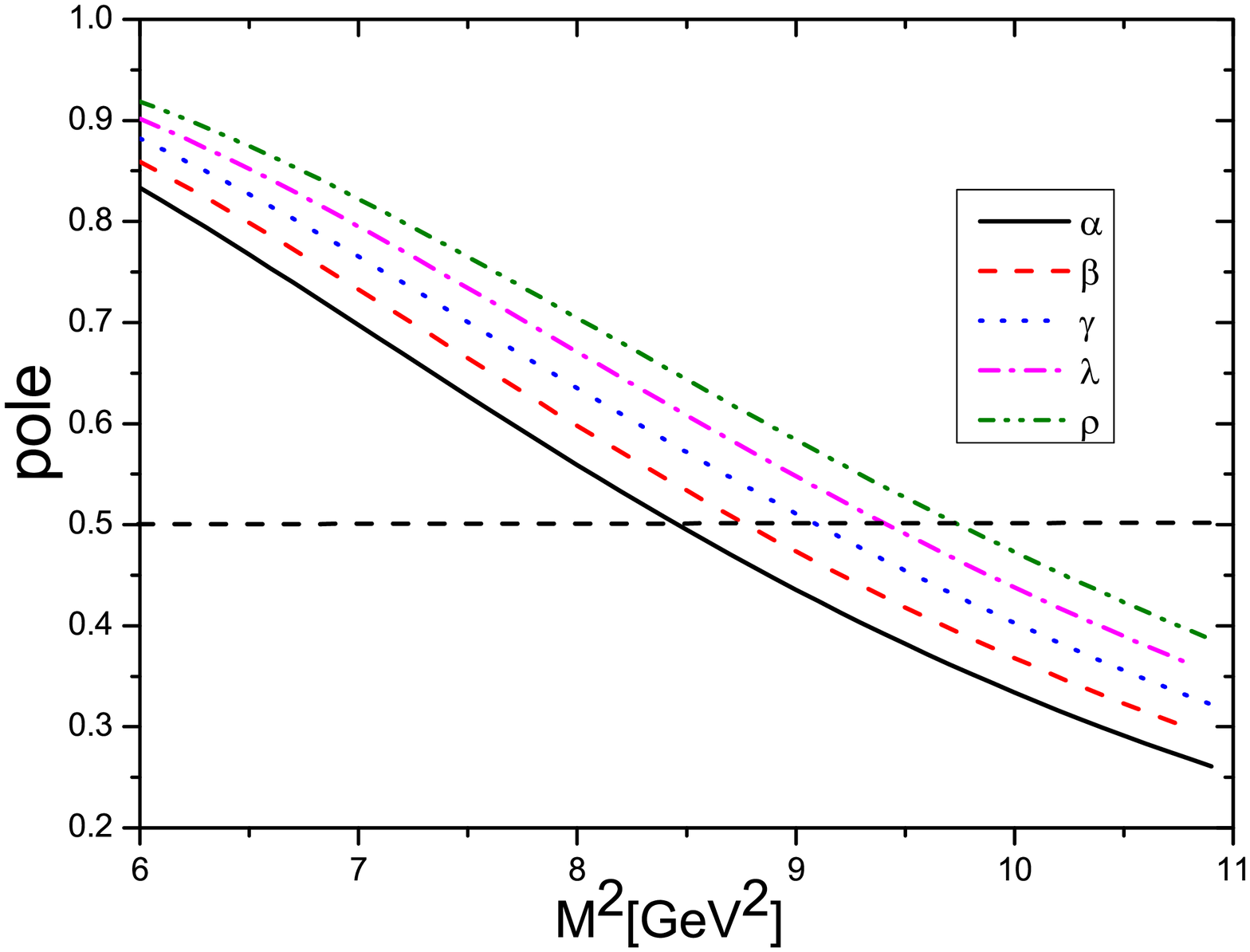}
\caption{The contributions from the pole terms with variation of the Borel parameter $M^2$. The notations $\alpha$, $\beta$, $\gamma$, $\lambda$ and $\rho$ correspond to the threshold parameters $s_0=121.00\,\rm{GeV}^2$, $123.21\,\rm{GeV}^2$, $125.44\,\rm{GeV}^2$, $127.69\,\rm{GeV}^2$ and $129.96\,\rm{GeV}^2$, respectively.}\label{fig8}
\end{figure}

\begin{figure}
\centering
\includegraphics[totalheight=5cm,width=7cm]{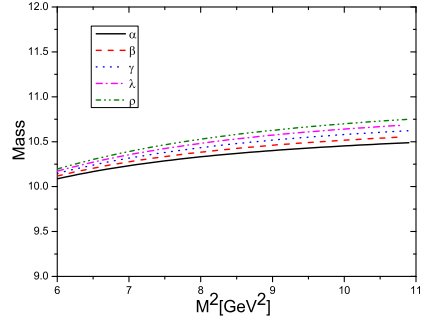}
\caption{The masses of the $B^*\bar{B}^*$ molecular state as
a function of $M^2$. The notations $\alpha$, $\beta$, $\gamma$, $\lambda$ and $\rho$ correspond to the threshold parameters $s_0=121.00\,\rm{GeV}^2$, $123.21\,\rm{GeV}^2$, $125.44\,\rm{GeV}^2$, $127.69\,\rm{GeV}^2$ and $129.96\,\rm{GeV}^2$.}\label{fig9}
\end{figure}

The contributions from the high dimension vacuum condensates in the OPE are shown in Fig.\ref{fig7}. We have used $s_0\geq 121.00\,\mbox{GeV}^2$ in the $B^{*}\bar{B^{*}}$ channel. From this figure it can be seen that for $M^2\geq 8.0\,\mbox{GeV}^2$ in the $B^{*}\bar{B^{*}}$ channel, the contribution of the dimension-$6$ condensate is less than $7\%$ of the total contribution and the contribution of the dimension-$5$ condensate is less than $18\%$ of the total contribution, which indicate a good OPE convergence. Therefore, we fix the uniform lower value of $M^2$ in the sum rule window as $M^2_{min}=8.0\,\mbox{GeV}^2$ in the $B^{*}\bar{B^{*}}$ channel. Fig.\ref{fig8} shows that the contributions from the pole terms with variation of the Borel parameter $M^2$ and the threshold parameters $s_0$. The upper limit $M^2_{max}$ is determined with the requirement that the pole term contribution is bigger than $50\%$ of the total contribution. We show in Table~\ref{tab:4} the values of $M^2_{max}$ for several values of $\sqrt{s_0}$ in the $B^{*}\bar{B^{*}}$. In Fig.\ref{fig9}, we show the molecular state masses, for different values of $\sqrt{s_0}$, in the relevant sum rule window. It can be seen that the mass is stable in the Borel window with the corresponding threshold $\sqrt{s_0}$.

Up to now we have fixed the values of the quark masses and condensates. To make the results more reliable, we also consider errors from the uncertainties of input parameters $m_{b}$, $\langle\bar{q}q\rangle$, and $m_{0}^{2}$. Taking into account both uncertainties of input parameters and uncertainties due to the continuum threshold parameter and Borel window, we finally arrive at
\begin{eqnarray}
M_{B^{*}\bar{B^{*}}} = (10.45\pm 0.31)~\mbox{GeV}.
\label{Zmass2}
\end{eqnarray}

\begin{table}
\caption{Upper limits in the Borel window for the axial-vector $B^*\bar{B}^*$ molecular current obtained from the sum rule for different values of $\sqrt{s_0}$.}\label{tab:4}
\begin{center}
\begin{tabular}{|c|c|}
\hline
$B^{*}\bar{B^{*}}$&$B^{*}\bar{B^{*}}$\\
$\sqrt{s_0}(\mbox{GeV})$&$M^2_{max}(\mbox{GeV}^2)$\\ \hline
11.0 &8.5 \\ \hline
11.1 &8.8 \\ \hline
11.2 &9.1 \\ \hline
11.3 &9.4 \\ \hline
11.4 &9.7 \\ \hline
\end{tabular}
\end{center}
\end{table}

\subsection{tetraquark state QCD sum rules}\label{sec3B}
As the case of $Z_{b}(10610)$, there are also two possible configurations for $Z_{b}(10650)$. In this subsection, we consider the tetraquark structure, which can be described by the interpolator:
\begin{eqnarray}\label{current2}
j^{\mu}(x)&=&\varepsilon^{\mu\nu\alpha\beta}{\epsilon_{abc}\epsilon_{dec}\over\sqrt{2}}(u(x)_a^T C\gamma_\nu b(x)_b)D_{\alpha}(\bar{d}(x)_d\gamma_\beta C\bar{b}(x)_e^T).
\end{eqnarray}

After some tedious calculations, the concrete forms of spectral densities can be derived:
\begin{eqnarray}
\rho^{OPE}(s)=\rho^{\mbox{pert}}(s)+\rho^{\langle\bar{q}q\rangle}(s)+\rho^{\langle
g^{2}G^{2}\rangle}(s)+\rho^{\langle
g\bar{q}\sigma\cdot G q\rangle}(s)+\rho^{\langle\bar{q}q\rangle^{2}}(s)+\rho^{\langle g^{3}G^{3}\rangle}(s),
\end{eqnarray}
with
\begin{eqnarray}\label{spectraltetra2}
\rho^{\mbox{pert}}(s)&=&\frac{1}{5*2^{9}\pi^{6}}\int_{\alpha_{min}}^{\alpha_{max}}\frac{d\alpha}{\alpha^{3}}\int_{\beta_{min}}^{1-\alpha}\frac{d\beta}{\beta^{4}}({\alpha}\beta+\beta^2-2{\alpha}-\beta)r(m_{b},s)^{5}
,\nonumber\\
\rho^{\langle\bar{q}q\rangle}(s)&=&\frac{\langle\bar{q}q\rangle}{3*2^{4}\pi^{4}}m_{b}\int_{\alpha_{min}}^{\alpha_{max}}\frac{d\alpha}{\alpha(1-\alpha)^3}[m_{b}^{2}-\alpha(1-\alpha)s]^3
\nonumber\\&&{}+\frac{\langle\bar{q}q\rangle}{3*2^{5}\pi^{4}}m_{b}\int_{\alpha_{min}}^{\alpha_{max}}\frac{d\alpha}{\alpha^{2}}\int_{\beta_{min}}^{1-\alpha}\frac{d\beta}{\beta^{3}}({\alpha}\beta+\beta^2-2{\alpha}-2\beta)r(m_{b},s)^{3}
,\nonumber\\
\rho^{\langle g^{2}G^{2}\rangle}(s)&=&\frac{\langle
g^{2}G^{2}\rangle}{9*2^{10}\pi^{6}}\int_{\alpha_{min}}^{\alpha_{max}}d\alpha\frac{(2\alpha+1)}{\alpha(\alpha-1)^3}[m_{b}^{2}-\alpha(1-\alpha)s]^3
\nonumber\\&&{}-\frac{\langle
g^{2}G^{2}\rangle}{9*2^{10}\pi^{6}}m_{b}^{4}\int_{\alpha_{min}}^{\alpha_{max}}\frac{d\alpha}{\alpha^4}\int_{\beta_{min}}^{1-\alpha}\frac{d\beta}{\beta^3}(1-\alpha-\beta)^2(5{\alpha}^4+2{\alpha}^3\beta
\nonumber\\&&{}-2{\alpha}^3+7{\alpha}\beta^3+4\beta^4+8\beta^3)r(m_{b},s)
\nonumber\\&&{}+\frac{\langle
g^{2}G^{2}\rangle}{9*2^{10}\pi^{6}}m_{b}^{2}\int_{\alpha_{min}}^{\alpha_{max}}\frac{d\alpha}{\alpha^4}\int_{\beta_{min}}^{1-\alpha}\frac{d\beta}{\beta^4}({\alpha}^6+3{\alpha}^5\beta-6{\alpha}^5-18{\alpha}^4\beta
\nonumber\\&&{}+21{\alpha}^4+10{\alpha}^3\beta^3-9{\alpha}^3\beta^2+15{\alpha}^3\beta-10{\alpha}^3+12{\alpha}^2\beta^4)r(m_{b},s)^{2}
\nonumber\\&&{}+\frac{\langle
g^{2}G^{2}\rangle}{9*2^{11}\pi^{6}}\int_{\alpha_{min}}^{\alpha_{max}}\frac{d\alpha}{\alpha^3}\int_{\beta_{min}}^{1-\alpha}\frac{d\beta}{\beta^3}(2{\alpha}^3+17{\alpha}^2-4{\alpha}\beta)r(m_{b},s)^3
,\nonumber\\
\rho^{\langle g\bar{q}\sigma\cdot G q\rangle}(s)&=&\frac{\langle
g\bar{q}\sigma\cdot G q\rangle}{2^{7}\pi^{4}}m_{b}\int_{\alpha_{min}}^{\alpha_{max}}d\alpha\frac{(7\alpha-5)}{\alpha(\alpha-1)^2}[m_{b}^{2}-\alpha(1-\alpha)s]^2
\nonumber\\&&{}+\frac{\langle
g\bar{q}\sigma\cdot G q\rangle}{2^{5}\pi^{4}}m_{b}\int_{\alpha_{min}}^{\alpha_{max}}\frac{d\alpha}{(\alpha-1)}{\alpha}s[m_{b}^{2}-\alpha(1-\alpha)s]
\nonumber\\&&{}
-\frac{\langle g\bar{s}\sigma\cdot
Gs\rangle}{2^{7}\pi^{4}}m_{b}\int_{\alpha_{min}}^{\alpha_{max}}\frac{d\alpha}{\alpha^2}\int_{\beta_{min}}^{1-\alpha}\frac{d\beta}{\beta^2}(\alpha^2+\beta^2-2\beta)r(m_{b},s)^2
,\nonumber\\
\rho^{\langle\bar{q}q\rangle^{2}}(s)&=&-\frac{\langle\bar{q}q\rangle^{2}}{9*2^{3}\pi^{2}}(8m_{b}^{4}+m_{b}^{2}s)\sqrt{1-4m_{b}^{2}/s}
,\nonumber\\
\rho^{\langle g^{3}G^{3}\rangle}(s)&=&-\frac{7\langle
g^{3}G^{3}\rangle}{9*2^{14}\pi^{6}}m_{b}^{2}\int_{\alpha_{min}}^{\alpha_{max}}d\alpha\int_{\beta_{min}}^{1-\alpha}\frac{d\beta}{\beta^{4}}(7\alpha^3+27\alpha^2\beta-39\alpha^2+21\alpha\beta^2
\nonumber\\&&{}-30{\alpha}\beta+9{\alpha}+\beta^3-3\beta^2+3\beta-1)r(m_{b},s)
\nonumber\\&&{}+\frac{\langle
g^{3}G^{3}\rangle}{3*2^{11}\pi^{6}}\int_{\alpha_{min}}^{\alpha_{max}}\frac{d\alpha}{\alpha}\int_{\beta_{min}}^{1-\alpha}\frac{d\beta}{\beta^{4}}(1-\alpha-\beta)^2r(m_{b},s)^2
\nonumber\\
\end{eqnarray}

It can be found that in the two configurations involved for $Z_{b}(10650)$, the dominant contributions to the sum rules are the perturbative term and $D=3$ condensate term, which are proportional by $3/4$ that will be canceled in the process of division for both cases. The deviation comes from the $D=4$ and $D=5$ condensate terms which play subdominant roles in the result. It tacitly suggests that the same conclusion can be drawn as in Sec.\ref{sec2} that two-point sum rules is unable to distinguish whether $Z_{b}(10650)$ is a molecular state or a tetraquark state. Our final numerical result is
\begin{eqnarray}
M_{[bd][\bar{b}\bar{u}]} = (10.48\pm 0.33)~\mbox{GeV}.
\label{Zmass3}
\end{eqnarray}
\section{Summary and conclusion}\label{sec4}
By assuming $Z_{b}(10610)$ as both a $B^{*}\bar{B}$ molecular state and a $[bd][\bar{b}\bar{u}]$ tetraquark state with quantum numbers $I^{G}J^{P}=1^{+}1^{+}$, the QCDSR approach has been applied to calculate the mass of the resonance. Our numerical results are $M_{Z}=(10.44\pm0.23)~\mbox{GeV}$ for molecular state and $M_{Z}=(10.50\pm0.19)~\mbox{GeV}$ for tetraquark state. Both of the results are compatible with the experimental data of $Z_{b}(10610)$ by Belle Collaboration. We also construct possible interpolators to describe the $Z_{b}(10650)$ as both an axial-vector $B^{*}\bar{B^{*}}$ molecular state and an axial-vector $[bd][\bar{b}\bar{u}]$ tetraquark state. Our numerical result are $M_{B^{*}\bar{B^{*}}} = (10.45\pm 0.31)~\mbox{GeV}$ and $m_{[bd][\bar{b}\bar{u}]} = (10.48\pm0.33)~\mbox{GeV}$, which are compatible with the experimental data of $Z_{b}(10650)$.

The calculations indicate that the mass sum rule could not distinguish $Z_{b}(10610)$ ($Z_{b}(10650)$) between a $B^{*}\bar{B}$ ($B^{*}\bar{B^{*}}$) molecular state and a $[bd][\bar{b}\bar{u}]$ tetraquark state. The clarification of the configuration of these two states requires further analysis on the decay channels $Z_{b}(10610)(Z_{b}(10650))\rightarrow \pi^{\pm}\Upsilon(nS)~~(n = 1, 2, 3)$ and $Z_{b}(10610)(Z_{b}(10650))\rightarrow\pi^{\pm}h_{b}(mP)~~(m =1, 2)$, which contain more detailed dynamical information.
\section*{Acknowledgement}
This work was supported in part by the National Natural Science
Foundation of China under Contract Nos.10975184 and 11047117, 11105222.

\end{document}